# Detection of magnetic gap in the topological surface states of MnBi$_2$Te$_4$

Haoran Ji[1], Yanzhao Liu[1], He Wang[2], Jiawei Luo[1], Jiaheng Li[3,4], Hao Li[5,6], Yang Wu[6,7], Yong Xu[3,4,8] and Jian Wang[1,3,9,10,*]

[1] International Center for Quantum Materials, School of Physics, Peking University, Beijing 100871, China

[2] Department of Physics, Capital Normal University, Beijing 100048, China

[3] State Key Laboratory of Low Dimensional Quantum Physics, Department of Physics, Tsinghua University, Beijing 100084, China

[4] Frontier Science Center for Quantum Information, Beijing, 100084, China

[5] School of Materials Science and Engineering, Tsinghua University, Beijing, 100084, China

[6] Tsinghua-Foxconn Nanotechnology Research Center and Department of Physics, Tsinghua University, Beijing 100084, China

[7] Department of Mechanical Engineering, Tsinghua University, Beijing, 100084, China

[8] RIKEN Center for Emergent Matter Science (CEMS), Wako, Saitama 351-0198, Japan

[9] CAS Center for Excellence in Topological Quantum Computation, University of Chinese Academy of Sciences, Beijing 100190, China

[10] Beijing Academy of Quantum Information Sciences, Beijing 100193, China

Abstract:

Recently, intrinsic antiferromagnetic topological insulator MnBi$_2$Te$_4$ has drawn intense research interest and leads to plenty of significant progress in physics and materials science by hosting quantum anomalous Hall effect, axion insulator state, and other quantum phases. An essential ingredient to realize these quantum states is the magnetic gap in the topological surface states induced by the out-of-plane ferromagnetism on the surface of MnBi$_2$Te$_4$. However, the experimental observations of the surface gap remain controversial. Here, we report the observation of the surface gap via the point contact tunneling spectroscopy. In agreement with theoretical calculations, the gap size is around 50 meV, which vanishes as the sample becomes paramagnetic with increasing temperature. The magnetoresistance hysteresis is detected through the point contact junction on the sample surface with an out-of-plane magnetic field, substantiating the surface ferromagnetism. Furthermore, the non-zero transport spin polarization coming from the ferromagnetism is determined by the point contact Andreev reflection spectroscopy. Combining these results, the magnetism-induced gap in topological surface states of $MnBi_2Te_4$ is revealed.

---

[*] Corresponding author. jianwangphysics@pku.edu.cn

Introducing ferromagnetism (FM) to a topological insulator (TI) can break the time-reversal symmetry and generate exotic quantum phases, leading to potential applications in the fields of spintronics and quantum computing [1-9]. A traditional method to realize a magnetic TI system is doping transition metal elements into TI thin films, such as Cr-doped $(Bi_{1-x}Sb_x)_2Te_3$ thin films [9]. However, the inevitable inhomogeneity produced in the doping process impedes the realization of a perfect magnetic TI. Hence, intrinsic magnetic TI single crystal is highly desired. Recently, the high-quality $MnBi_2Te_4$ single crystal was successfully synthesized and verified to be such an intrinsic magnetic TI [10-15]. $MnBi_2Te_4$ and its family have rapidly drawn broad interest as promising platforms to investigate many long-sought topological phenomena. For example, quantum anomalous Hall effect, axion insulator state, and high-Chern-number insulator phase have been reported in $MnBi_2Te_4$ thin film devices [13-15]. Currently, one of the research interests focuses on the experimental detection of the magnetic gap in the topological surface states of $MnBi_2Te_4$, which is theoretically predicted to be induced by surface FM [16,17]. So far, several angle-resolved photoemission spectroscopy (ARPES) experiments have been carried out to detect the surface gap. One ARPES work reports a surface gap even at temperatures above the antiferromagnetism (AFM) transition temperature (Néel temperature $T_N \sim 24$ K) [10]. However, some other ARPES works suggest that the topological surface states remain gapless across the $T_N$ [18-20]. Thus, the ARPES results of the gapped surface states are controversial and inconsistent with the theoretical predictions. Further evidence, especially via different experimental techniques, are strongly desired.

The needle-anvil point contact spectroscopy method is a powerful tool to study the local properties of the sample surface by measuring the electrical transport through a small constriction [21-25]. Two types of point contact spectroscopy are adopted to study the sample properties: point contact tunneling spectroscopy (PCTS) [22,23] and point contact Andreev reflection spectroscopy (PCARS) [24,25]. The PCTS method is carried out with a relatively large contact barrier at the tip-sample interface, in which the electron tunneling is dominant in the electrical transport across the contact. In the PCTS measurements, the obtained point contact spectra (PCS) provide the information of the electronic density of states (DOS) around the Fermi level in the sample surface [26]. The PCARS method is usually performed on the superconducting point contact formed by a superconductor and a metal. In the PCARS measurements, the transport spin polarization of the ferromagnetic sample can be measured [25,27,28].

In this letter, high-quality $MnBi_2Te_4$ single crystals are characterized by systematic transport and magnetization measurements. The electronic and magnetic properties of $MnBi_2Te_4$ are consistent with previous reports [10,11,14]. The electronic structure on the (111) surface of $MnBi_2Te_4$ is studied through PCTS measurements. The obtained PCS characterize an energy gap, which is around 50 meV at 2 K and disappears around $T_N \sim 24$ K. The gap size and temperature-dependent evolution are consistent with theoretical predictions on the magnetic gap in topological surface states. The out-of-plane magnetoresistance hysteresis is obtained through point contact junctions on the sample surface, showing the surface FM in $MnBi_2Te_4$. The transport spin polarization of approximately 0.5 is determined by the PCARS measurements, further supporting the surface FM. The combined results suggest the experimental detection of the magnetism-induced gap in topological surface states [16,17].

$MnBi_2Te_4$ is a van der Waals layered material and consists of Te-Bi-Te-Mn-Te-Bi-Te septuple layers (SLs) stacking in the ABC sequence along the out-of-plane direction [12,16], as shown in the inset of Fig. 1(a). It exhibits A-type AFM in the bulk and two-dimensional (2D) FM within each SL [10-12,29]. The electronic and magnetic properties of $MnBi_2Te_4$ single crystals are characterized by systematic transport and magnetization measurements. Figures 1(a)-1(c) display the electrical transport results. The standard six-electrode method was applied for the measurements, as shown in the inset of Fig. 1(b). The temperature dependence of longitudinal resistivity ($\rho_{xx}$-$T$) is shown in Fig. 1(a). A distinct kink at around 24 K suggests the Néel temperature of our sample is about 24 K, consistent with previous reports [12,14]. Figures 1(b) and 1(c) show the longitudinal resistivity $\rho_{xx}$ and the Hall resistivity ($\rho_{yx}$) as a function of the out-of-plane magnetic field at different temperatures. Below the Néel temperature, two transitions can be observed in both magnetoresistance and Hall trace curves with increasing magnetic field (Figs. 1(b) and 1(c)). The corresponding critical fields are marked as $B_{c1}$ and $B_{c2}$, respectively. The lower critical field $B_{c1}$ is about 3.3 T at 1.8 K. At $B_{c1}$, both $\rho_{xx}$ and $\rho_{yx}$ exhibit one remarkable drop. Previous studies attribute the drop at $B_{c1}$ to the occurrence of the spin-flop and the system entering a metastable canted AFM (cAFM) state [12,14,30]. As the magnetic field further increases and exceeds $B_{c2}$ (~ 7 T), weak anomalies are detected in $\rho_{xx}$ and $\rho_{yx}$, indicating that the spins are fully polarized and the system is driven into a FM state. These two critical fields $B_{c1}$ and $B_{c2}$ become lower as the temperature increases and vanishes above $T_N$. The negative slopes of the Hall traces indicate that the charge carriers of bulk are electron-type. At the low field region below $B_{c1}$, the $\rho_{yx}$ shown in Fig. 1(c) linearly decreases with increasing magnetic field. The carrier density and the mobility at 1.8 K are $1.08 \times 10^{20} \text{cm}^{-3}$ and $70 \text{cm}^2/\text{Vs}$, respectively, close to the previous report [14].

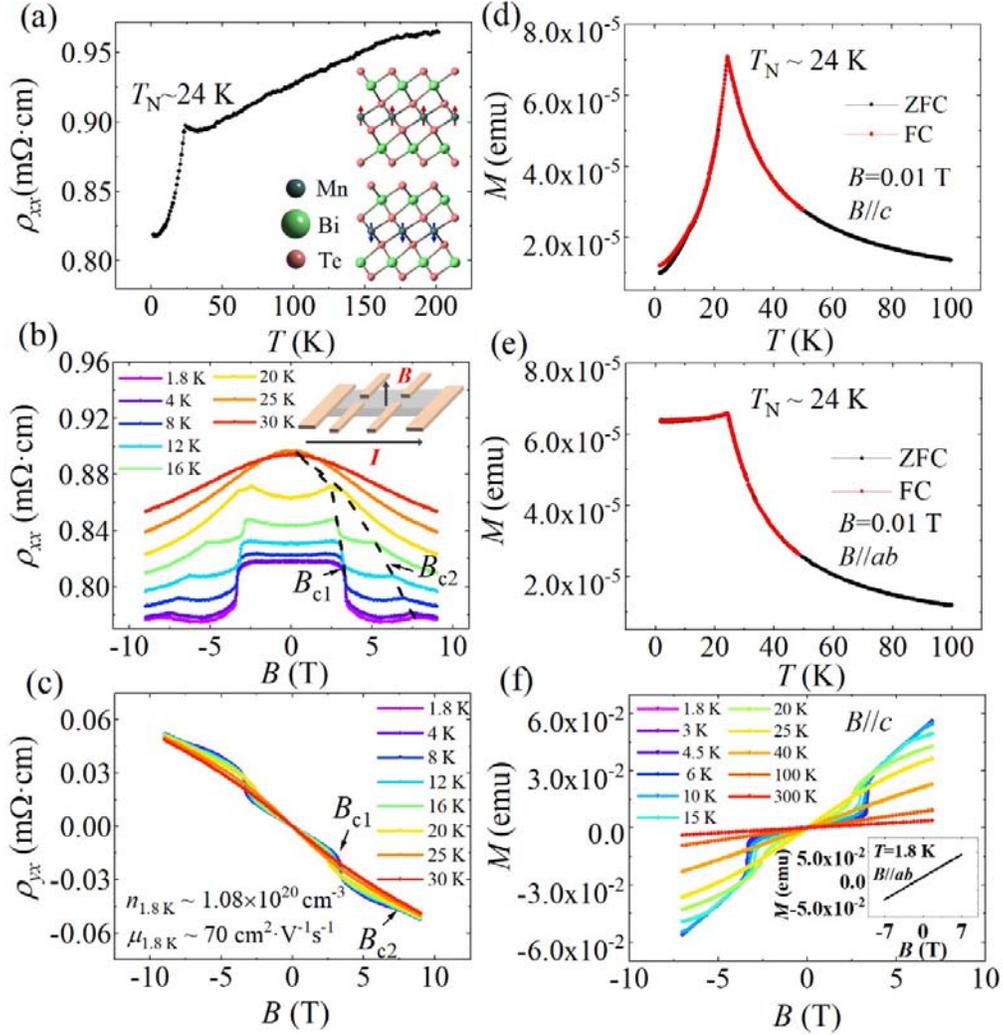

FIG. 1. Transport and magnetic properties of MnBi$_2$Te$_4$ single crystal. (a) Temperature dependence of $\rho_{xx}$ measured from 200 K to 1.8 K. The kink around 24 K reveals the AFM transition. The inset shows the crystal structure of MnBi$_2$Te$_4$. Magnetic moment directions of Mn ions are denoted by red and blue arrows. (b)-(c) $\rho_{xx}$ and $\rho_{yx}$ at different temperatures in out-of-plane magnetic field. The lower and upper magnetic fields are labeled by $B_{c1}$ and $B_{c2}$, respectively. Inset of (b) is a schematic structure of the electrical transport measurements. (d)-(e) $M$-$T$ curves of MnBi$_2$Te$_4$ measured under ZFC and FC process in a magnetic field of 0.01 T applied out-of-plane (d) and in-plane (e), respectively. (f) Magnetization versus out-of-plane magnetic field at different temperatures. Inset of (f) shows the in-plane magnetic field dependence of magnetization at 1.8 K.

The magnetization measurement results are shown in Figs. 1(d)-1(f). Figures 1(d) and 1(e) display the temperature dependence of magnetization (*M-T*) measured in zero-field cooling (ZFC) and field cooling (FC) process under a magnetic field of 0.01 T. Magnetic field is applied along and perpendicular to the *c* axis, respectively. A kink around 24 K reveals the AFM transition, consistent with the transport measurement results. Curves of magnetization (*M*) as a function of out-of-plane magnetic field (*B*) at different temperatures are plotted in Fig. 1(f). At 1.8 K, the critical field of AFM to cAFM transition ($B_{c1}$) is around 3.3 T and $B_{c1}$ disappears above $T_N$. On the contrary, only linear field dependence can be observed when the magnetic field is applied in the *ab* plane, as shown in the inset of Fig. 1(f). The spin-flop transition in *M-H* curves and anisotropic magnetization agree with the predicted A-type AFM feature [11,12].

To study the surface states and surface FM of $MnBi_2Te_4$, we performed point contact experiments with a standard needle-anvil configuration (inset of Fig. 2(a)). The point contact is established between a mechanically sharpened metallic or superconducting tip and the (111) surface of $MnBi_2Te_4$ mounted on Attocube piezo-positioners. Differential conductance (d*I*/d*V*) of the PCS is obtained by standard lock-in technique in quasi-four-probe configuration. The experiments are conducted in a dilution refrigerator from Leiden.

Firstly, the PCTS measurements using platinum-iridium (PtIr) tips are performed to probe the local density of states (DOS) of the sample surface [26]. In our experiments, the sample is intentionally exposed to atmosphere, not only to create an oxidized barrier at the tip-sample contact interface [23], but also to modulate the surface Fermi level of the sample [14,18]. The d*I*/d*V* spectra of PCTS measurements exhibit a common feature: the linear conductance structure (LCS) in a bias voltage range over tens of millivolts, which is a signature of the tunneling type point contact [31- 36]. Typical PCS at different temperatures are shown in Fig. 2(a), where the d*I*/d*V* value exhibits a linear increase with applied bias voltage, leading to the LCS in the spectra. This kind of LCS in the spectra has been reported in previous tunneling junctions and PCTS experiments with an oxidized or anodized surface [34-36]. The LCS in the d*I*/d*V* spectra is normally ascribed to the inelastic tunneling effect at the contact interface [31-35]. As seen from the two d*I*/d*V* curves in Fig. 2(a), at 6 K, there is a sharp minimum structure around zero bias voltage, which is smoothed out as the temperature increases to 24 K. This thermal smearing effect is consistent with the inelastic tunneling scenario, which can be characterized by the full width at half maximum (FWHM) of the second derivative of differential conductance $d^2G/dV^2$, where $G$ denotes the $dI/dV$ [31]. As shown in Fig. 2(b), the FWHM of $d^2G/dV^2$ of the spectra at different temperatures exhibit a monotonic and linear increase with a slope of 6.1 $k_BT$ ($k_B$ is the Boltzmann constant). This temperature-dependent behavior is in agreement with the theoretical calculations [31]. Thus, the LCS in the measured PCS indicates the corresponding point contacts are in the tunneling limit and meet the PCTS conditions. More discussions on the PCTS measurements can be found in the Supplemental Material [37]. Since the LCS is the inelastic tunneling feature originating from the tunneling barrier [31-36], it does not affect the intrinsic PCTS features from the sample [32].

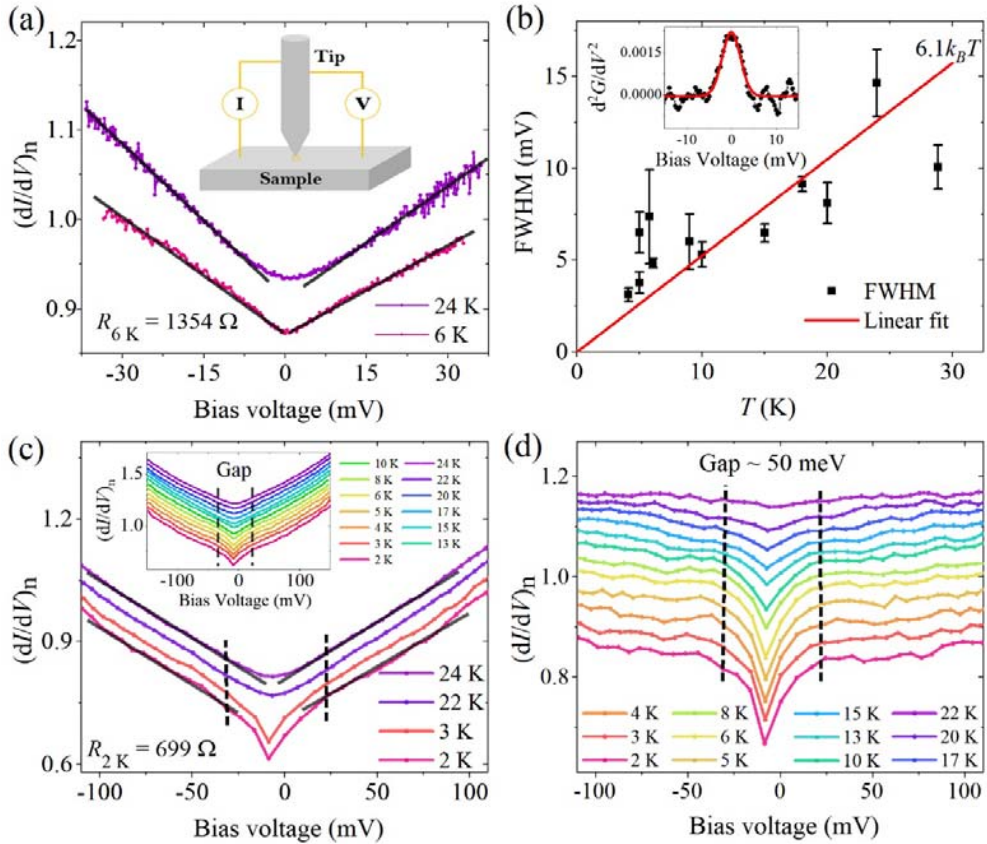

FIG. 2. The PCTS measurements on MnBi$_2$Te$_4$ single crystal with a PtIr tip. (a) A typical type of normalized $dI/dV$ spectra obtained at 6 K and 24 K that show a large V-shape LCS. Solid lines are guiding lines for the LCS. Inset in (a) is a schematic of the point contact configuration. (b) The FWHM of the $d^2G/dV^2$ of the spectra and the best linear fit. The dots line in the inset in (b) is the $d^2G/dV^2$ around zero bias of the spectrum at 6 K in (a). The red solid line is the best Gauss fitting curve that gives the FWHM value of the $d^2G/dV^2$. (c) Normalized $dI/dV$ spectra at selected temperatures from another point contact state. Two vertical dash lines indicate the detected dip feature that deviates from the background. Solid lines are guiding lines for the LCS. The spectra are shifted for clarity. Inset in (c) is the whole set of normalized and shifted spectra of (c). Two vertical dash lines indicate the detected gap. (d) Normalized $dI/dV$ spectra after removal of the background from (c). Two vertical dash lines indicate the detected dip feature. The spectra are shifted for clarity.

In addition to the LCS, spectral features implying a gapped electronic structure are also observed on multiple point contact states. One typical set of PCS is shown in the inset of Fig. 2(c). The spectra at selected temperatures are shown in Fig. 2(c) for clarity. All the spectra exhibit the distinct LCS, indicating the point contact is in the tunneling limit. The main difference in Fig.2(c), compared with the PCS in Fig. 2(a), is that a zero bias conductance dip feature emerges at relatively low temperature and vanishes at high temperature. For the PCS at 2 K and 3 K shown in Fig. 2(c), it can be clearly distinguished that a sharp dip structure emerges within several tens of millivolt, deviating from the LCS. Here, the spectrum at 24 K is taken as a background and subtracted from this set of PCTS result, since the dip structure completely disappears at 24 K and no clear difference can be found when temperature increases from 22 K to 24 K. The subtracted curves are shown in Fig. 2(d) where the dip structure gets more distinct. The characteristic energy around 50 meV can be determined by the two kinks in the PCS. With increasing temperature, the dip structure gradually smears out and disappears above 22 K. The conductance dip in the $dI/dV$ spectra with such temperature-dependence is generally taken as a sign of an energy gap [47-49]. In our experiment, the gap completely disappears at 24 K, which is close to the Néel temperature of our $MnBi_2Te_4$ sample (see Fig. 1). Thus, the occurrence of the gap structure is likely connected to the magnetism of $MnBi_2Te_4$. Also, the characteristic energy scale of 50 meV is consistent with the theoretical predictions of the magnetic gap in the topological surface states on the (111) surface of $MnBi_2Te_4$ [16]. Given the characteristics of the observed gap and its consistencies with theoretical calculations, the gap structure in PCTS curves can be interpreted as the magnetism-induced gap in the topological surface states.

The gap is also detected in other point contact states with similar critical temperature and energy size (see Supplemental Material [37]). However, the gap structure is absent in most point contact states. Some possible mechanisms may contribute to this phenomenon, including the large energy difference between the Fermi level and the location of the surface gap, the imperfect magnetic order on the surface [18,20], etc. The small chance of the detection of the gap indicates that only small parts of the sample surface host the gapped surface states. Such an assumption may explain the gapless surface states results observed in previous ARPES works, considering that ARPES results are the spatially averaged signals collected in the light spot. In our point contact measurements, sharp tips with needle sizes smaller than a few micrometers are used, which makes it possible to detect the surface gap in some small areas of the sample surface.

The magnetic origin of the gap is implied by its occurrence around $T_N$. According to the theoretical prediction [16], the spins ferromagnetically couple in each SL, and an out-of-plane FM emerges on the surface of $MnBi_2Te_4$ below $T_N$. The surface FM, once exists, may be detected through surface sensitive techniques, like point contact measurements. Figure 3 shows the magnetoresistance of the point contacts on the (111) surface of $MnBi_2Te_4$ using a Nb tip. The magnetoresistance (MR) is defined by $MR = \frac{R(B)-R(0)}{R(0)} \times 100\%$, and $R$ denotes the differential resistance in point contact measurements. The hysteretic behavior with two sharp dips at $B_\perp \sim \pm 0.7$ T is observed by sweeping the out-of-plane magnetic field (Fig. 3(a)). In contrast, no hysteresis is observed by sweeping the in-plane magnetic field (Fig. 3(b)). The obtained MR hysteresis substantiates the FM on the $MnBi_2Te_4$ surface, which orients along the c-axis, supporting the magnetic origin of the gap on the (111) surface of $MnBi_2Te_4$.

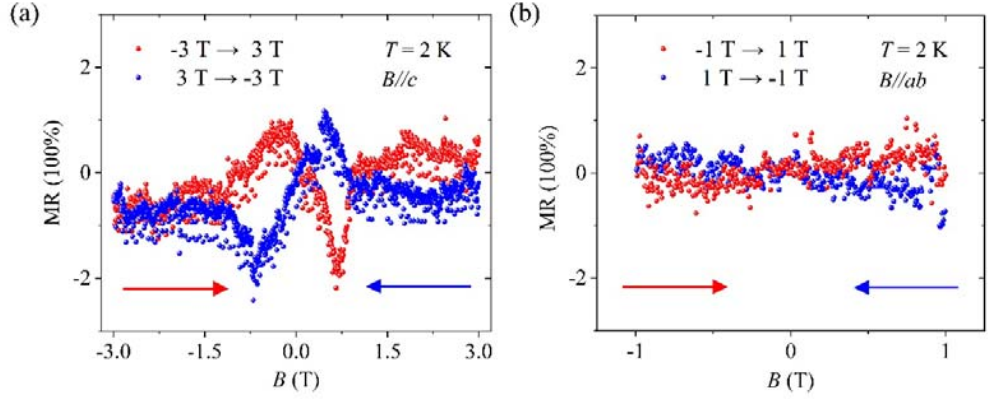

FIG. 3. Magnetoresistance of $MnBi_2Te_4$ single crystal obtained by point contact measurement using a Nb tip at 2 K with (a) an out-of-plane magnetic field and (b) an in-plane magnetic field.

To further demonstrate the surface FM of $MnBi_2Te_4$, we carried out the PCARS measurements by pressing superconducting niobium (Nb) tips on the (111) surface of $MnBi_2Te_4$ to form a superconducting point contact [25,27,28]. In a ferromagnet, the conduction electrons are spin-polarized, resulting in a difference in spin population near the Fermi level. If a point contact is formed between a ferromagnetic sample and a conventional superconductor, the Andreev reflection process would be limited by the spin minority electrons around the Fermi level in the process of forming spin-singlet cooper pairs. The current through the contact can be viewed as two parts, namely the un-polarized current that obeys the conventional BTK theory and the fully-polarized current that is entirely a quasiparticle current. Therefore, the resultant spin polarization coming from the ferromagnetic sample surface can be quantitatively studied by fitting the d$I$/d$V$ curve with the BTK model incorporating spin-polarization [25,27,28].

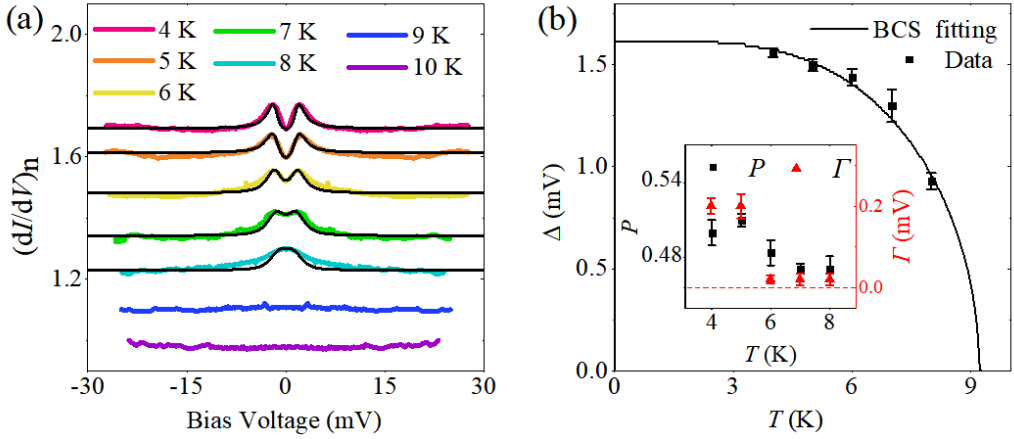

FIG. 4. The PCARS measurement on MnBi$_2$Te$_4$ single crystal with a superconducting Nb tip. (a) Normalized d$I$/d$V$ spectra at selected temperatures. The spectra are symmetrized for the convenience of fitting. The solid lines are the theoretical spin-polarized BTK fitting. The curves are shifted for clarity. (b)The hollow dots are the Δ values obtained from the spin-polarized BTK fitting. The solid line is the BCS fitting curve, giving a BCS ratio of 2Δ/$k_B T_c$=4.04. Inset in (b) shows the fitted spin-polarization parameter $P$ and broadening parameter $\varGamma$ at different temperatures.

Here, the PCARS measurement is conducted below 10 K, well below the Néel temperature of MnBi$_2$Te$_4$ of 24 K. As shown in Fig. 4(a), the symmetrized spectra of the PCARS measurement show a clear double-peak structure that fades away as temperature gradually reaching 9 K, which is close to the superconducting transition temperature of Nb ($T_c$ = 9.26 K) [60]. The double-peak structure and its temperature-dependent behaviors are in accord with the features of superconducting PCS in the ballistic regime [21,25]. The superconductivity conductance enhancement in the detected PCS is less than 10% and the d$I$/d$V$ minimum around zero bias reaches a value that is close to the value of the normal state. These kinds of spectra are possibly due to two mechanisms: one is a relatively large interface barrier strength (parameter $Z$ in the modified BTK model) with proper broadening effects (parameter $\varGamma$) and the other is the spin-polarization (parameter $P$ in the spin-polarized BTK model) existing in the sample surface. In the former condition, the PCS can be fitted with the BTK model without spin-polarization.

However, in our work, three unreasonable fitting results make this condition improper: i) the parameter $Z$ cannot be fixed in the fitting process; ii) the BCS fitting gives a $T_c$ of 9.9 K, which is larger than the $T_c$ of Nb; iii) the parameter $\Gamma$ is too large to stabilize a superconducting phase. The BTK fitting results without spin-polarization and the detailed discussions can be found in the Supplemental Material [37]. In the latter condition, the modified BTK model incorporating spin-polarization is applied. The fitting curves of the experimental data are shown in Fig. 4(a). The fitting result gives a temperature-independent $Z = 0.04$ and the $\Gamma$ values that are far smaller than the $\Delta$ values. According to the fitting results, the superconducting gap vs temperature behavior agrees with the BCS characteristic (Fig. 4(b)) and the BCS fitting values of $\Delta = 1.61$ meV ($T = 0$ K) and $T_c = 9.26$ K are consistent with the properties of Nb ($\Delta=1.53$ meV, $T_c = 9.26$ K) [60]. The spin-polarization is around 50% and changes little with increasing temperature, as shown in the inset of Fig. 4(b). Thus, the spin-polarized BTK model can well explain the PCARS results, and the spin-polarization from the out-of-plane electron transport supports the surface FM of $MnBi_2Te_4$.

In summary, we have investigated the surface states of $MnBi_2Te_4$ by point contact measurements. A surface gap of $MnBi_2Te_4$ emerging just below the Néel temperature is detected via PCTS measurements. The gap exhibits a characteristic energy scale of ~50 meV, and gradually smears with increasing temperature and eventually disappears around 24 K. These properties are consistent with the theoretical prediction of the magnetism-induced gap in the topological surface states in $MnBi_2Te_4$. Furthermore, the out-of-plane MR hysteresis is observed through the point contacts on the surface of $MnBi_2Te_4$, substantiating the surface FM. Additionally, the out-of-plane transport spin polarization on the surface is determined through the PCARS measurement, which further demonstrates the surface FM of $MnBi_2Te_4$. Our research clarifies an essential property of the magnetic TI $MnBi_2Te_4$: the magnetism-induced gap in the topological surface states, which is the foundation for understanding many novel topological phenomena and quantum phenomena in this material, such as topological magneto-electric or magneto-optic effects.

We thank Yanan Li for helpful discussions. This work was supported by National Natural

**Supplementary Material for**

**Detection of magnetic gap in the topological surface states of MnBi$_2$Te$_4$**

Haoran Ji[1], Yanzhao Liu[1], He Wang[2], Jiawei Luo[1], Jiaheng Li[3,4], Hao Li[5,6], Yang Wu[6,7], Yong Xu[3,4,8] and Jian Wang[1,3,9,10]

[1]International Center for Quantum Materials, School of Physics, Peking University, Beijing 100871, China

[2]Department of Physics, Capital Normal University, Beijing 100048, China

[3]State Key Laboratory of Low Dimensional Quantum Physics, Department of Physics, Tsinghua University, Beijing 100084, China

[4]Frontier Science Center for Quantum Information, Beijing, 100084, China

[5]School of Materials Science and Engineering, Tsinghua University, Beijing, 100084, China

[6]Tsinghua-Foxconn Nanotechnology Research Center and Department of Physics, Tsinghua University, Beijing 100084, China

[7]Department of Mechanical Engineering, Tsinghua University, Beijing, 100084, China

[8]RIKEN Center for Emergent Matter Science (CEMS), Wako, Saitama 351-0198, Japan

[9]CAS Center for Excellence in Topological Quantum Computation, University of Chinese Academy of Sciences, Beijing 100190, China

[10]Beijing Academy of Quantum Information Sciences, Beijing 100193, China

I. Repeatable observations of the surface gap on multiple PC states through PCTS.

The point contact spectra (PCS) showing the surface gap structure, as well as the linear conductance structure (LCS), are also obtained on other PC states. Fig. S1 (a) and (b) are obtained from one PC state and Fig. S1(c) and (d) are from another PC state, both using a niobium (Nb) tip. In the first PC state, the surface gap is approximately 40 meV and emerges below 24 K, shown in Fig. S1(a) and (b). In the second state, the surface states gap is about 40 meV and emerges around 18 K, shown in Fig. S1(c) and (d). The energy sizes and the critical temperatures of these two results are consistent with the properties of the gap shown in Fig. 2 in the main text, confirming the reproducibility of our observation.

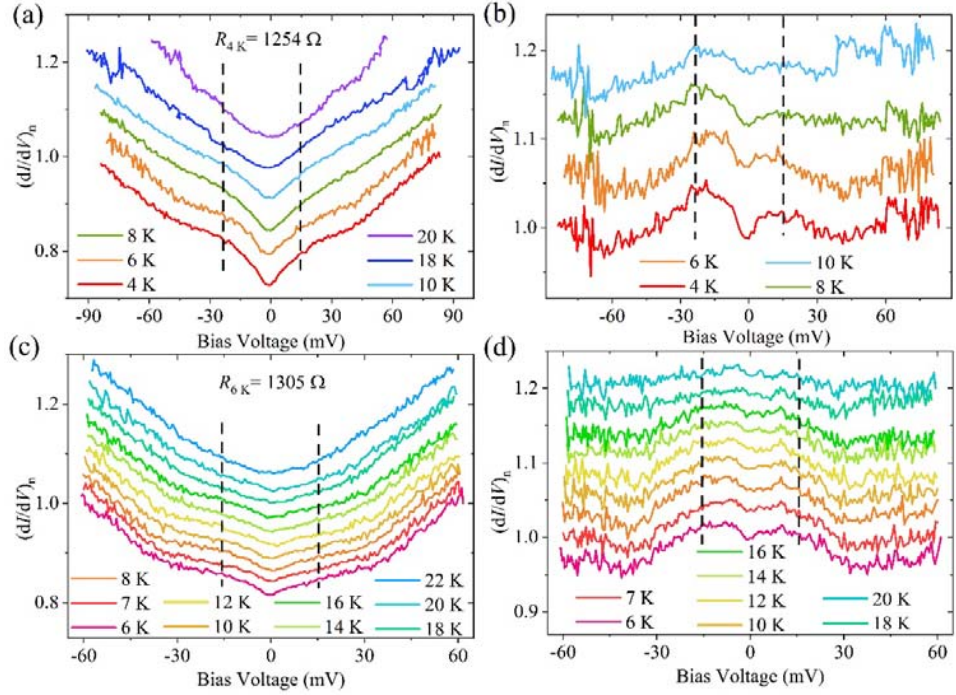

Fig. S1. Two PC states of the PCTS measurements on $MnBi_2Te_4$ single crystal with a Nb tip. (a) Normalized d$I$/d$V$ spectra at selected temperatures of one PC state. The spectra are shifted for clarity. Two vertical dash lines indicate the detected gap. The zero bias contact resistance is 1254 Ω. (b) Normalized d$I$/d$V$ spectra after removal of the background from (a). The spectra are shifted for clarity. Two vertical dash lines indicate the gap. (c) Normalized d$I$/d$V$ spectra at selected temperatures of another PC state. The spectra are shifted for clarity. Two vertical dash lines indicate the gap. (d) Normalized d$I$/d$V$ spectra after removal of the background from (c). The spectra are shifted for clarity. Two vertical dash lines indicate the gap. The zero bias contact resistance is 1305 Ω.

II. Theoretical analysis for point contact Andreev reflection spectroscopy (PCARS) data shown in Fig. 4 of the main text

Fig. S2(a) represents the original PCS of Fig. 4 in the main text that has not been symmetrized. The double-peak structure is evident at low temperatures and the d$I$/d$V$ curves become flat above the superconducting transition temperature of Nb ($T_c$ = 9.26 K). As shown in Fig. S2(b), the modified Blonder-Tinkham-Klapwijk (BTK) model without spin-polarization is applied for fitting. The corresponding parameters are shown in Fig. S2(c)). The hollow dots are the superconducting energy gap Δ obtained from the BTK fitting process. The BCS fitting is also applied to the temperature-dependent value of Δ, shown in Fig. S2(c)). The BCS fitting results give Δ = 1.38 meV ($T$ = 0 K) and $T_c$ = 9.9 K, obviously contradicting with the properties of Nb (Δ=1.53 meV, $T_c$=9.26 K). The inset of Fig. S2(c)) shows the temperature-dependent values of the broadening parameter $\Gamma$ (red) and the interface barrier strength $Z$ (black). The maximum and minimum values of the parameter $Z$ are 0.88 meV and 0.54 meV respectively. The large variation range of $Z$ value is unreasonable since the parameter $Z$ is supposed to be temperature-independent in the BTK fitting process. Most importantly, the ratio of $\Gamma/\Delta$ is around 80%, which is too large to be acceptable for BTK fitting results. Because the parameter $\Gamma$ is related to the quasiparticle scattering rate and this parameter value should be smaller than superconducting gap Δ for the stabilization of the superconducting phase. Thus, the possibility that the typical PCS shown in Fig. S2(b)) are mainly contributed by the large $Z$ without $P$ is excluded.

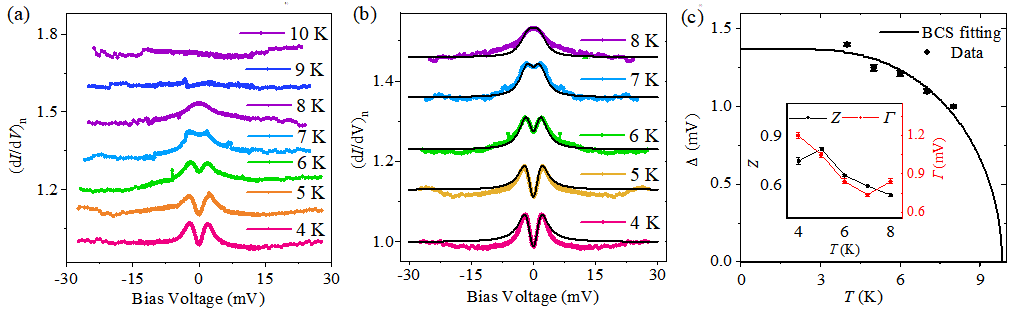

Fig. S2 The PCARS results and the conventional BTK fitting results of the PC state shown in Fig. 4 in the main text. (a) Normalized spectra at selected temperatures without symmetrizing process. (b) The symmetrized d$I$/d$V$ spectra of (a). The solid lines are the best BTK fitting results without considering spin-polarization. The curves are shifted for clarity. (c) The hollow dots are the Δ values obtained from the conventional BTK fitting without spin-polarization. The solid line is the BCS fitting curve, giving a BCS ratio of $2\Delta/k_BT_c$=3.24. Inset in (c) shows the parameter $Z$ and parameter $\Gamma$ at different temperatures.

III. Spin polarized BTK fitting curves without shift.

Selected PCARS data and corresponding spin-polarized fitting curves from Fig. 4(a) are shown in Fig. S3 without shifts. The double-peak structure is well fitted, suggesting the good quality of fitting with the spin-polarized BTK model.

Theoretical BTK models are based on an ideal, ballistic, and one-dimensional point contact [1], which may sometimes fail to describe the real condition comprehensively. For example, the critical current dips outside the superconducting double peaks, as one typical phenomenon in point contact measurements, cannot be fitted by present theoretical models [2]. Therefore, the traditional treatment of the fitting process is mainly focused on the double-peak feature and ignores the dips [2,3]. Considering this fitting criterion, the spin-polarized BTK model can well explain the experimental curves, since the double-peak structure can be well fitted.

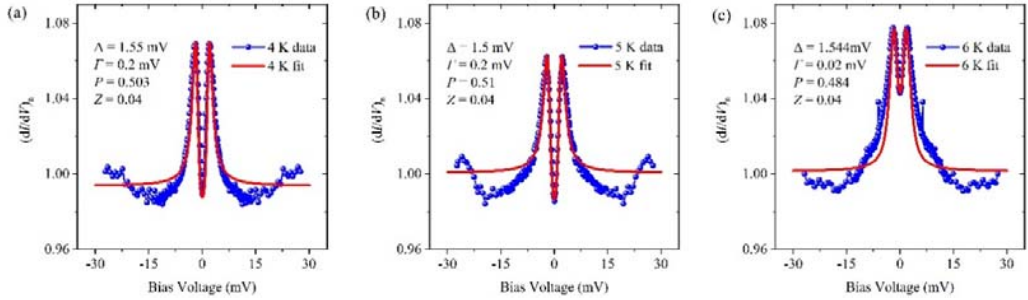

Fig. S3 Selected normalized spectra at (a) 4 K, (b) 5 K, and (c) 6 K from the data shown in Fig. 4. The solid lines are the spin-polarized BTK fitting curves of Fig. 4(a). The best-fit parameters are also shown in the figures.

IV. Comparison in the $\Gamma$ and $\Delta$ behaviors between the spin polarized BTK fitting and the conventional BTK fitting.

The $\Gamma$ parameter is the broadening term in the conventional BTK model and spin-polarized BTK model, related to the shortening of the quasiparticle lifetime. Once the $\Gamma$ is too large, the lifetime of quasiparticles is too short for a stable superconducting phase and the superconductivity is expected to be substantially suppressed in the PC region in principle. Therefore, the $\Gamma$ value is required to be much smaller than the energy gap ($\Delta$).

In our spin-polarized BTK fitting result, the $\Gamma$ value is 0.2 meV from 4 K to 5 K, and the value is 0.02 meV from 6 K to 8 K. The largest ratio of $\Gamma$ divided by $\Delta$ about is 13.3% and the smallest ratio is only 1.4% (shown in Fig. S4(a)). The $\Gamma$ value is much smaller than the $\Delta$ value. Such fitting values are reasonable results for a ballistic PC measurement, which indicates that the electrons can transport through the contact with nearly no scattering. Small $\Gamma$ values that are close to zero, or are exactly zero, have been reported in both conventional BTK fitting and spin polarized BTK fitting of ballistic PC measurement [1-7].

By contrast, the largest ratio of $\Gamma$ divided by $\Delta$ about 85.7% and the smallest ratio is still 66.9% in the fitting result of the conventional BTK model without spin polarization (shown in Fig. S4(b)). The $\Gamma$ parameter and the $\Delta$ value at each temperature have nearly the same magnitude of order.

The superconductivity conductance enhancement (compared to the conductance of normal state) in Fig. 4 is less than 10%, which is a general spectroscopic feature of PCARS with spin polarization [1,5]. The imbalance between spin-up and spin-down electrons suppresses the process of Andreev reflection and, consequently, suppress the superconducting enhancement in the PCS. In PC measurement without spin polarization, the conductance enhancements with a reasonably small $\Gamma$ parameter are generally over 20% and sometimes are over 60% [6,7]. Therefore, conventional BTK fitting without spin polarization cannot explain our PCARS results.

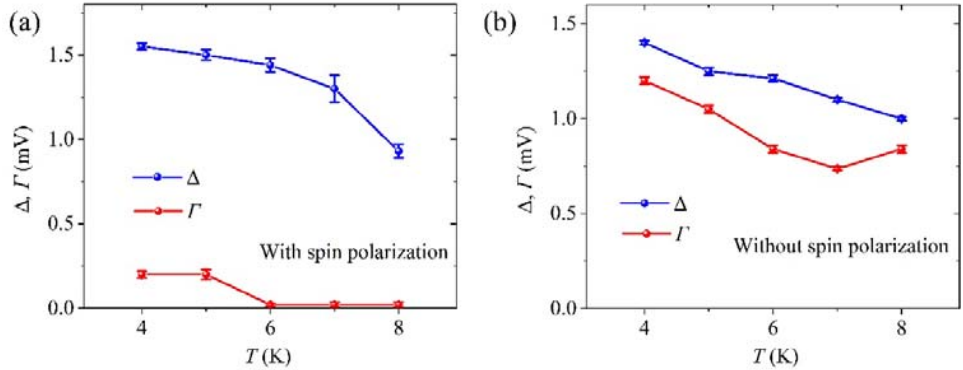

Fig. S4 Temperature dependence of the superconducting energy gap ($\Delta$), and the broadening parameter ($\Gamma$) obtained through (a) the spin-polarized BTK fitting results of Fig. 4, and (b) conventional BTK model fitting results of Fig. S2.

V. Comparison between our work and previous ARPES reports and theoretical works on the magnetic gap in the topological surface state of $MnBi_2Te_4$.

Before our work, mainly ARPES measurements were carried out to detect the surface gap of the intrinsic topological insulator $MnBi_2Te_4$. The mostly probed surface is the (111) surface of $MnBi_2Te_4$, which is predicted to host gapped topological surface states [23]. A detailed comparison to the published results is shown in Table. 1. So far, only our work is consistent with the theoretical predictions in both the emergency temperature and the gap size.

|  | Surface gap | Method | Note |
|---|---|---|---|
| Our work | ~50 meV (2 K) | PCTS measurement | The gap closes around 24 K |
| *Nature* **576**, 416 (2019) | ~70 meV (17 K) | ARPES measurement | The gap is still present at 35 K |
| *Nature* **576**, 416 (2019) | 88 meV | Theoretical calculation |  |
| *Phys. Rev. Research* **1**, 012011 (2019) | ~85 meV (5 K) | ARPES measurement | The gap is still around 85 meV at 300 K |
| *Sci. Adv.* **5**: eaaw5685 (2019) | ~50 meV | Theoretical calculation |  |
| *Phys. Rev. Lett.* **122**, 206401 (2019) | ~50 meV | Theoretical calculation |  |
| *Phys. Rev. X* **9**, 041038 (2019) | Gapless | ARPES measurement |  |
| *Phys. Rev. X* **9**, 041039 (2019) | Gapless | ARPES measurement |  |
| *Phys. Rev. X* **9**, 041040 (2019) | Gapless | ARPES measurement |  |
| *Phys. Rev. Lett.* **125**, 117205 (2020) | Gapless | ARPES measurement |  |

Table. 1 Comparison between our work and published works on the gap structure of the topological surface states.

VI. Discussions about the tunneling regime point contact measurement.

The point contact tunneling spectroscopy (PCTS) refers to the spectroscopy measured in point contact (PC) junction with a relatively large contact barrier between sample and tip [8,44]. In the PCTS measurement, the electron tunneling is dominant in the electrical transport across the contact. Hence, through measuring the differential conductance (d$I$/d$V$) spectra, PCTS can provide the electronic structure features of the detected samples [10]. Electronic gaps, like superconducting energy gap, charge density wave gap, and Kondo gap, can be studied by PCTS [11-14].

First, in our measurement, an insulating oxide barrier is intentionally formed in the tip-sample interface by exposure to the atmosphere. It is a common treatment to take the native oxidized surfaces as the insulating barrier in tunneling measurements [12,15]. Correspondingly, our point contact resistances are very large (699 Ω for Fig. 2(c) at 2 K, 1254 Ω for Fig. S1 (a) at 4 K, and 1305 Ω for Fig. S1 (c) at 6 K). These results are consistent with the scenario of the tunneling point contact, whose resistances are usually from hundreds of ohms to thousands of ohms [8,16], and are subsequently larger than typical metallic point contact whose resistances are about several ohms to tens of ohms [17].

Second, LCS in a wide bias voltage range is observed in nearly all PCTS spectra. Normally, the LCS is ascribed to the inelastic tunneling process and is a signature of tunneling spectra [18-21]. For comparison, such LCS is absent in metallic PC [17]. In the inelastic tunneling process of PCTS, electrons tunnel through the barrier and scatter with the broad and flat distribution of energy-loss modes in the barrier, leading to the asymmetric and LCS in the d$I$/d$V$ spectra. The corresponding LCS follows(d$I$/d$V$) $\propto \int_0^{eV} F(E)dE$, where e$V$ is the electron energy and $F(E) \approx$ constant is the spectral distribution of inelastic modes in the barrier [19,22]. The LCS in the spectra indicates that the data are obtained in the tunneling regime.

Additionally, the thermal smearing behavior of our LCS is further consistent with the inelastic tunneling scenario, which further confirms the tunneling is the main conduction mechanism in our PCTS measurements. As seen from the two d$I$/d$V$ curves in Fig. 2(a) in the main text, at 6 K, there is a sharp minimum structure around zero bias voltage, which is smoothed out as the temperature increases to 24 K. This result is consistent with previous reports on tunneling measurements, where the sharp minimum structure of the LCS gradually smears out with increasing temperature [19]. The second derivative of conductance d$^2G$/d$V^2$ around zero bias is generally used to characterize this smearing. As shown in Fig. 2(b) in the main text, the $d^2G/dV^2$ around zero bias of our PCTS spectra exhibits a monotonic and linear temperature-dependence with a slope of approximately 6.1 $k_BT$ ($k_B$ is the Boltzmann constant). This temperature-dependent behavior is in agreement with the theoretical calculation and experimental work on tunneling measurements of 6 $k_BT$ [19]. Thus, the LCS in the measured PCS indicates that our PCTS measurements are in the tunneling regime.

VII. Discussions about the origin of the gap structure in the PCTS measurement.

Two decisive features of the gap structure in the PCTS measurement are that the gap emerges just below $T_N$ (~ 24 K) and exhibits a gap size of 50 meV, in agreement with the theoretical prediction [23].

According to the theoretical calculation, the ferromagnetism forms in each septuple layer of MnBi$_2$Te$_4$ and exhibits antiferromagnetic coupling along the c-axis below $T_N$. The resultant out-of-plane ferromagnetism in the surface layer of MnBi$_2$Te$_4$ induces a gap in the topological surface states [23]. The gap structure in our spectra, emerging just below the antiferromagnetic (AFM) - paramagnetic (PM) transition and diminishing as temperature increasing to $T_N$, accords with such theoretical prediction. Additionally, the transport spin-polarization and the magnetoresistance hysteresis obtained through PC measurements indicate the out-of-plane ferromagnetism in the surface and support the magnetism-induced gap in MnBi$_2$Te$_4$.

Excitations, like phonon, may also lead to spectra with some kink structures. Phonon mechanism, for example, does not agree with our spectra for the following reasons. Firstly, multiple kinks are generally observed in the point contact measurement as there are always multiple phonon modes [3,25], which is at odds with our observations. Secondly, in point contact measurement, the phonon-induced features are generally studied through the second derivative of conductance $d^2I/dV^2$ spectra because the intensity of electron-phonon interaction is too small to be distinguished in $dI/dV$ spectra [3,25], which contradicts with the clear and distinguishable structure in our $dI/dV$ spectra. Additionally, the characteristic energy scale of the gap structure about 50 meV and the emergency just below $T_N$ are consistent with the theoretical predictions about the surface gap [23,26].

Point contact spectroscopy, as a surface sensitive measurement, mainly obtains the properties of the contact region on the sample surface [28,30]. As a magnetic topological insulator, MnBi$_2$Te$_4$ has a bulk bandgap around 200 meV, which does not change across the AFM transition [23,27,28]. Although contributions from bulk states cannot be fully excluded in PC measurement in principle, the 50 meV gap structure observed in PCTS measurements below $T_N$ does not agree with the bulk bandgap. The consistencies with the predicted surface gap in energy size and magnetic origin can exclude other possibilities, including excitations that may not disappear around $T_N$, and the bulk gap which is as large as 200 meV [23,27,28], etc. The combined results mentioned above indicate that a magnetic gap in the topological surface states is detected in our PCTS measurements.

VIII. Brief introduction about the spin polarized BTK model

Point contact measurement is a well-established probe for quantitatively measuring spin polarization in ferromagnetic materials. In the superconducting point contact with a ferromagnetic sample, the imbalance in the number of spin-up and spin-down electrons at the Fermi level of the sample will limit the Andreev reflection. The transport spin polarization quantity P can be defined as

$$P = \frac{N_\uparrow(E_F)v_{F\uparrow} - N_\downarrow(E_F)v_{F\downarrow}}{N_\uparrow(E_F)v_{F\uparrow} + N_\downarrow(E_F)v_{F\downarrow}},$$

where $N_\uparrow E_F$ and $N_\downarrow E_F$ are the spin-dependent density of states and $v_{F\uparrow}$ and $v_{F\downarrow}$ are corresponding Fermi velocity. Then, the total current can be divided into two parts: $I=(1-P)I_u+PI_P$. $I_P$ ($I_u$) denotes fully polarized (un-polarized) current. For paramagnetic metal, $P=0$; and for half metal, $P=1$. The unpolarized current, $I_u$, carries no net spin polarization and obeys the conventional BTK theory. The remaining current, $I_P$, is entirely a quasiparticle current and can be calculated by allowing only non-Andreev processes at the point contact. More information can be found in the reference [29].

IX. Discussions about the LCS on different contact conditions.

The term "linear conductance structure" specifically refers to the spectroscopic feature, which is normally attributed to the inelastic tunneling effect in the barrier [19,21]. It usually exhibits a linear shape in a large bias voltage range, but may deviate from a straight line to a certain degree [31,33]. Also, the LCS may be asymmetric with respect to zero bias voltage.

The deviation from linearity at relatively high bias voltage has been reported previously [31]. A quadratic modification term ($|V|V$, where $V$ is applied bias voltage) due to the contact condition was theoretically proposed to describe this deviation in the LCS [33]. And small differences in the spectra can be considered as a result of the different barrier conditions [22,34,35]. In our measurement, the small deviation from linearity appears over 80 meV (Fig. 2(c)) and is well above the gap size, which may consequently have a negligible effect on the gap structure. However, with this deviation from linearity in mind, we actually take the spectrum at the highest temperature in measurement as a background rather than a straight line to subtract the possible deviation from linearity. For example, in Fig. 2(c), the curve at 24 K (around $T_N$) is chosen as the background for subtraction to further analyze the gap structure at low temperatures.

The slight asymmetry is also present in PCS, which may partially be ascribed to the asymmetry of the LCS. The asymmetry of the spectra is a common feature in PCTS measurements, which has also been reported in previous works [13,35,36]. Theoretical explanations have been proposed and successful in fitting experimental data by considering the position of the inelastic scattering center [31,33] in the barrier and the ratio of Fermi energy to the parameter of the tunneling barrier strength [33]. Therefore, the asymmetry of the PCS depends on the contact barrier, and the asymmetry of the spectra may vary from different point contact states.